\documentclass[twocolumn,showpacs,preprintnumbers,amsmath,amssymb]{revtex4}

\usepackage{graphicx}
\usepackage{dcolumn}
\usepackage{bm}

\begin{document}


\title{Contactless estimation of critical current density in thin superconducting films and its temperature dependence using ac magnetic measurements}

\author{Ahmed Youssef}
\affiliation{Charles University in Prague, Faculty of Mathematics and Physics, Ke Karlovu 3, 121 16 Prague 2,
Czech Republic}
\author{Lucia Bani\v{c}ov\'{a}}
\affiliation{Charles University in Prague, Faculty of Mathematics and Physics, Ke Karlovu 3, 121 16 Prague 2,
Czech Republic}
\author{Zden\v{e}k \v{S}vindrych}
\affiliation{Institute of Physics of the AS CR, v.v.i., Na Slovance 2, 182 21, Prague, Czech Republic}
\author{Zden\v{e}k Jan\accent23 u}
\affiliation{Institute of Physics of the AS CR, v.v.i., Na Slovance 2, 182 21, Prague, Czech Republic}
\email{janu@fzu.cz}

\date{\today}

\begin{abstract}

We have measured magnetic response of second-generation high temperature superconductor YBa$_2$Cu$_3$O$_x$ wire and Nb thin film in perpendicular ac field as a function of temperature. We compare experimental complex ac susceptibility to the calculated susceptibility based on the model of Bean's critical-state response in two-dimensional (2D) disk in perpendicular field. The harmonic analysis is needed for comparison between the model and the experimental data. We present a method of linking model and experimental susceptibility. Good agreement of experimental susceptibility with model susceptibility of 2D disk allows contactless estimation of critical depinning current density and its temperature dependence.

\end{abstract}

\maketitle

Superconducting thin films are widely used in superconducting electronics and power applications. Nb films are an essential part of magnetic flux sensors (SQUID), radiation detectors or resonance cavities. A magnetic flux penetrating into the film may deteriorate performance of these devices \cite{Brandt93}. A second generation of high temperature superconducting wires for power applications is based on RE-Ba$_2$Cu$_3$O$_x$ (YBCO) thin films. Recently, materials with strong flux pinning carried critical depinning current densities exceeding hundreds of GA/m$^{2}$ at zero or liquid helium temperature, one or two orders higher values than were achieved in BSCCO round wires or in MgB$_2$, Nb-Ti, Nb$_3$Sn, and Nb$_3$Al \cite{FSU}. Unlike BSCCO wires whose performance is lowered by a flux flow at temperature above 35 K the YBCO wires operate even at liquid nitrogen temperature.

Understanding of physical processes related to vortex dynamics in the thin films and their models are important for development and applications. Complete analytical models of a response in time varying field are known only for restricted geometries and current-voltage relation $E(j)$. The models exist only for an infinitely long cylinder and slab in parallel field and disk and strip in perpendicular field \cite{Brandt1997}. The $E(j)$ is limited to a step-like dependence of a quasistatic Bean model with the critical depinning current density $j_{c}$. Until the current density is $|j| < j_{c}$, the electric field is $E = 0$. When $j = \pm j_{c}$ the electric field appears and flux density (vortices) rearranges to reduce a flux gradient to the critical value. For other geometries or nonlinear $E(j)$ relation (flux diffusion and flow) the model has to be solved numerically, typically using finite element method.

The model to 2D disk in perpendicular applied time varying field was worked out by Clem and Sanchez \cite{Clem1994}. We have shown that this model correctly predicts the response of the YBCO wire and allows estimate in a contactless way the critical current density and its temperature dependence \cite{Youssef09}. In this paper we reanalyze earlier measurements on Nb film \cite{Youssef08} using a method first applied to the YBCO wire and we compare the results.

The Nb film of thickness of 250 nm was deposited by a dc magnetron sputtering in Ar gas on 400 nm thick silicon-dioxide buffer layer which was grown by a thermal oxidation of a silicon single crystal wafer \cite{May1999}. The film is poly-crystalline with texture of a preferred orientation in the (110) direction and is highly tensile. Grain size is about 100 nm. The square samples of 5 $\times$ 5 mm$^2$ in dimensions were cut out from the 3-inch wafer.

Second-generation high temperature superconductor wire (2G
HTS wire) \cite{SuperPower2008A} consists of a 50 $\mu$m nonmagnetic nickel alloy substrate (Hastelloy), 0.2 $\mu$m of a textured MgO-based buffer stack deposited by an assisting ion beam, 1 $\mu$m RE-Ba$_2$Cu$_3$O$_x$ superconducting layer SmYBaCuO deposited by metallo-organic chemical vapor deposition, and 2 $\mu$m of Ag, with 40 $\mu$m total thickness of surround copper stabilizer (20 $\mu$m each side) \cite{SuperPower2008,SuperPower2008B}. The sample is cut into 4 mm long segment of 4 mm wide wire.

Measurements were performed in a noncommercial continuous operating SQUID magnetometer. An applied homogeneous field is generated by a superconducting solenoid and may be of essentially arbitrary waveform. An immobile sample is placed in one coil of a gradiometer coupled to the SQUID. Magnetization loops are recoded at fixed ac field amplitude $H_{ac}$ and frequency $f$ upon cooling or warming. The sample temperature sensor is GaAlAs diode and control temperature sensor is Si diode. A signal proportional to the magnetic moment of the sample, $m(t)$, and signal of the applied field $h(t)$ (proportional to a supply current) are digitized and
recorded on a hard disk. The complex ac susceptibility $\chi (f) = M(f)$ / $H(f)$ is calculated in real time using discrete fast Fourier transform \cite{Youssef09}. The $M(f)$ and $H(f)$ are Fourier transforms of $m(t)$ and $h(t)$, respectively.

The model susceptibility is calculated in the same way from model magnetization curves \cite{Youssef09}.
We assume temperature dependence of the critical depinning current density

\begin{equation}
\label{jc}
\frac{j_{c}(T)}{j_{c}(0)} =\frac{H_{d}(T)}{H_{d}(0)} = \left[ 1 - \left( \frac{T}{T_{c}} \right)^{m} \right]^{n},
\end{equation}

\noindent where $H_{d} = d j_{c} / 2$ is the characteristic field and $d$ is the film thickness. Model temperature is obtained using the inverse function for Eq. (\ref{jc}) and multiplying both the numerator and denumerator, $H_{d}(T) / H_{d}(0)$, by $H_{ac}$. The model and experimental susceptibilities are matched using four free parameters, $c \equiv H_{ac} / H_{d}(0) $, $m$, $n$, and $T_{c}$,

\begin{equation}
\label{map}
\left\{ \left[ 1 - \left( \frac{c H_{d}}{H_{ac}} \right)^{\frac{1}{n}} \right]^{\frac{1}{m}}, \chi \left(\frac{H_{d}}{H_{ac}} \right) \right\} \leftrightarrow \left\{ \frac{T}{T_{c}}, \chi \left( T \right) \right\}.
\end{equation}

\noindent Zero temperature critical depinning current density is $j_{c}(0) = 2 H_{ac} / cd$.

\begin{figure}
\includegraphics[scale = 0.5]{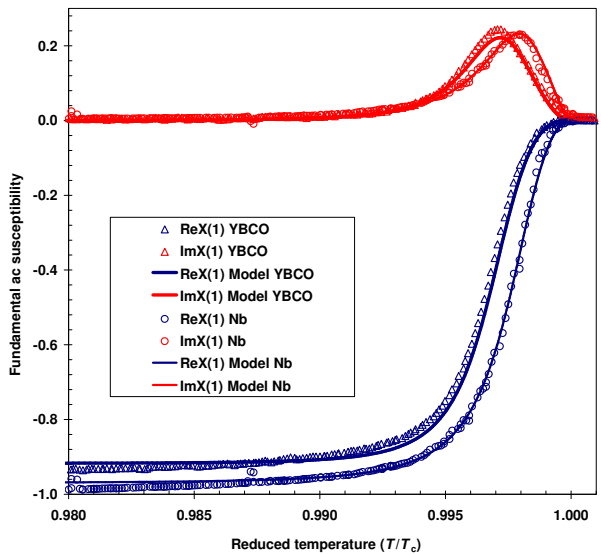}
\caption{\label{fig1} Fundamental ac susceptibility as a function of temperature at $\mu_{0}H_{ac}$ = 10 $\mu$T and frequency of 1.5625 Hz. The marks are experimental susceptibilities and curves are model susceptibilities.}
\end{figure}

\begin{figure}
\includegraphics[scale = 0.5]{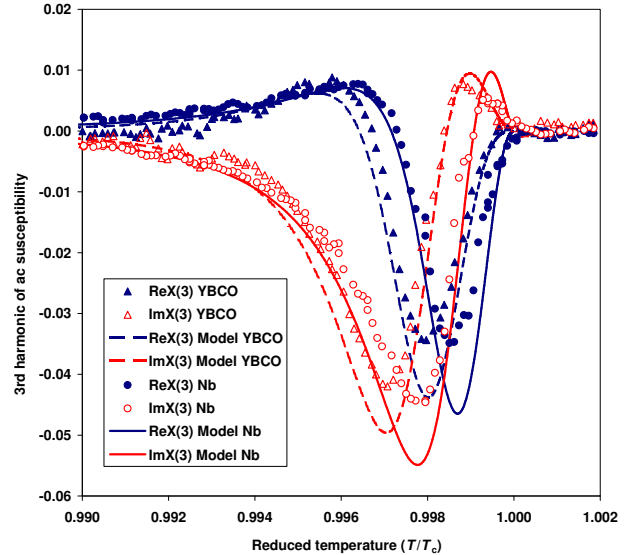}
\caption{\label{fig2} Third harmonic of ac susceptibility as a function of temperature at $\mu_{0}H_{ac}$ = 10 $\mu$T and frequency of 1.5625 Hz. The marks are experimental susceptibilities and curves are model susceptibilities.}
\end{figure}

Experiments were performed in zero dc field. Temperature dependence of the experimental and model fundamental ac susceptibility is shown in Fig.~\ref{fig1}. Nb film was measured at cooling rate 0.1 K/min and YBCO wire at cooling rate 1 K/min. The model susceptibility is vertically scaled by a factor $s$ to fit particularly the section of the experimental susceptibility $\chi (T)$ that is related to the critical state. Evidently, the temperature dependent real part of $\chi (T)$ for $T/T_{c} < 0.9$ is not related to the critical state. For Nb $s = 0.97$ and the model susceptibilities are plotted against $1-(cH_{d}/H_{ac})^{2/3}$. For YBCO $s = 0.92$ and the model susceptibilities are plotted against $1-(cH_{d}/H_{ac})^{1/2}$. The fundamental ac susceptibility alone is not sufficient to find unambiguously the parameters $n$, $m$, $c$, and $T_{c}$. One needs also match the third harmonic of the ac susceptibility, see Fig.~\ref{fig2}.

In conclusion, using the contactless magnetic measurements we have found the critical depinning current density in Nb film $j_{c}(0)= 3 \times 10^{11}$ A/m$^{2}$ with temperature dependence $j_{c}(T) = j_{c}(0) [ 1 - (T/T_{c})]^{3/2}$ in comparison with $j_{c}(0)= 10^{12}$ A/m$^{2}$ and steeper temperature dependence $j_{c}(T) = j_{c}(0) [ 1 - (T/T_{c})]^{2}$ found for the YBCO wire. In both cases the exponent falls into the typical range from 1.5 to 3.5 for the temperature dependence of a pinning force.

This work was supported by Institutional Research Plan Contract No. AVOZ10100520, grant ESO MNT-ERA (ME10069 of MEYS CR), Research Project MSM Contract No. 0021620834, SVV-2010-261303, and ESF program NES.

{}

\end{document}